\documentclass{article}

\usepackage{arxiv}

\usepackage{amssymb,amsmath}
\usepackage{graphicx,epsfig}
\usepackage{url}
\usepackage{algorithm,algorithmic}
\usepackage{subfigure}
\usepackage{xcolor,colortbl}

\begin{document}

\title{Query Term Weighting based on Query Performance Prediction}

\author{
  Haggai Roitman\\
  IBM Research\\
  Haifa University Campus\\
  Haifa, Israel \\
  \texttt{haggai@il.ibm.com} \\
  }

\date{}

\maketitle

\begin{abstract}
This work presents a general query term weighting approach based on query performance prediction (QPP). To this end, a given term is weighed according to its predicted effect on query performance. Such an effect is assumed to be manifested in the responses made by the underlying retrieval method for the original query and its (simple) variants in the form of a single-term expanded query. Focusing on search re-ranking as the underlying application, the effectiveness of the proposed term weighting approach is demonstrated using several state-of-the-art QPP methods evaluated over TREC corpora. 
\end{abstract}

\section{Introduction}\label{sec:intro}
This work presents a simple and general query term weighting approach based on query performance prediction (QPP)~\cite{CarmelBook}. To this end, a given term is weighed according to its predicted effect on query performance. 
Such an effect is assumed to be manifested in the responses made  by the underlying retrieval method for the original query and its (simple) variants. Query variants are introduced in the form of a single-term expanded query, which artificially ``re-focus" the intent of the original query. Hence, for a given term, the query is expanded with that single term and is resubmitted to obtain the corresponding response. By evaluating the quality of both result lists (i.e., the one originally retrieved for the query and the one retrieved for its single-term expanded version), the marginal effect of that term on the query's performance may be estimated. A post-retrieval QPP method~\cite{CarmelBook}, assumed to be provided as an input, serves in this work as a proxy for estimating such ``before"  and ``after" effects on query performance.
The relative difference (positive, negative or none) in predicted query performance is, therefore, used for determining a given term's importance with respect to the original query.

Weighed terms may be extracted from any source, e.g., either considering the terms explicitly specified in the query or implicitly derived from it (e.g., using relevance models~\cite{Lavrenko:2001}). Specifically, in this work, terms are derived from the RM3 pseudo relevance model~\cite{Lavrenko:2001}.
Focusing on search re-ranking as the underlying application, the effectiveness of the proposed term weighting approach is demonstrated using several state-of-the-art QPP methods evaluated over TREC corpora. The proposed term weighting approach is further demonstrated to provide a more robust retrieval by improving the performance of the underlying relevance model that is used to derive the terms.

\section{Related Work}\label{sec:related}
There exist numerous approaches for term weighting in IR, spanning from more ``traditional" unsupervised term weighting schemes (e.g., TF-IDF, BM25~\cite{Robertson:2009}, language models~\cite{Ponte:1998}, etc) and relevance models~\cite{Lavrenko:2001}, to supervised methods that exploit various term features (e.g., local and global term statistics~\cite{Cao:2008,Bendersky:2008}, term dependencies or proximity~\cite{Bendersky:2010,Svore:2010,Bendersky:2012a}, external corpora or sources~\cite{Bendersky:2011,Bendersky:2012}, etc).

In this work, term weights are derived using query performance prediction (QPP) methods. In the absence of any prior relevance knowledge, QPP methods may be utilized for predicting retrieval effectiveness~\cite{CarmelBook}. Many QPP methods have been proposed, including various pre-retrieval methods, post-retrieval methods and their combinations~\cite{CarmelBook}.

The proposed approach shares some relationship with previous works that utilized QPP methods for ranker selection~\cite{liu2006experiments,Balasubramanian:2010a,Roitman:2014}, selective query expansion~\cite{amati2004,cronen2004language,diaz2016pseudo,Roitman:2018:SYR} and query reduction~\cite{Balasubramanian:2010b}.
Yet, unlike such previous works, QPP methods are used in this work to \textbf{directly weight terms for document scoring}, rather than just for choosing over retrieved lists or selecting query terms for expansion/reduction.

Several previous works have also utilized various QPP methods for term weighting~\cite{Bendersky:2008,Song:2011,Ozdemiray:2014}. Yet, this work proposed term weighting approach is different from such previous works in several ways. First, similar to Song et al.~\cite{Song:2011} and unlike the works by Bendersky and Croft~\cite{Bendersky:2008} and Ozdemiray and Altingovde~\cite{Ozdemiray:2014}, term weighting is determined in the scope of the original query.  In~\cite{Bendersky:2008,Ozdemiray:2014} term weights are independently derived from the original query by estimating the performance of a query that contains only a single term. Second, unlike Song et al.~\cite{Song:2011}, the proposed weighting approach may be used to weight terms that were not explicitly specified in the query (e.g., those terms that are derived from a relevance model). Finally, this work proposed term weighting approach provides a more holistic solution, where various QPP methods may be applied, including methods similar to those that were proposed in~\cite{Bendersky:2008,Song:2011,Ozdemiray:2014}.

\section{Approach}\label{sec:solution}

\subsection{Preliminaries}
Let $q$ denote a query submitted over a corpus of documents $\mathcal{D}$.
Given query $q$, let $Score_{q}(d)$ denote the score of document $d\in\mathcal{D}$ and let $D_{q}^{[k]}\subseteq{\mathcal{D}}$ further denote a ranked list that contains the $k$ highest scored documents according to $Score_{q}(d)$. In this work, the retrieval scores $Score_{q}(d)$ are calculated according to the \emph{query-likelihood} model~\cite{Ponte:1998} as follows.
Let $w$ be a term in the vocabulary $V$ and $p_{x}^{[\mu]}(w)$ be text $x$'s Dirichlet smoothed language model with smoothing parameter $\mu$~\cite{Ponte:1998}, calculated as follows:
\begin{equation}
 p^{[\mu]}_x(w)=\frac{tf(w,x)+\mu\frac{tf(w,\mathcal{D})}{|\mathcal{D}|}}{|x|+\mu},
\end{equation}
where $tf(w,\cdot)$ and $|\cdot|$ denote the term frequency of term $w$ and the overall term frequency (either in text $x$ or the entire collection $\mathcal{D}$), respectively.
Let $\{q_i\}$ be the bag of query terms, the query-likelihood  of a document $d\in\mathcal{D}$ is calculated as follows:
\begin{equation}\label{ql model}
Score_{q}(d)= p^{[\mu]}_d(q)= \log \prod_{q_i} p^{[\mu]}_d(q_i).
\end{equation}

\subsection{Term Weighting using QPP}

For a given query $q$ and a term $w\in{V}$, let $q\vee{w}$ denote the expansion of query $q$ with $w$ as a single additional disjunctive term. Such an expansion basically ``shifts" the original information need expressed in $q$ towards a specific term $w$. Surely, for a given query $q$, not all terms $w\in{V}$ may be related to $q$. Hence, an expansion of such terms may result in a query-drift~\cite{Buckley:2004}. To minimize query drift risk, terms $w\in{V}$ considered for such an expansion should be carefully selected. For example, in this work we make use of the RM3 relevance model~\cite{Lavrenko:2001} to derive terms $w$ that are (presumably) related to $q$. 

Terms $w\in{V}$ are now weighted according to their predicted effect on the performance of query $q$. While a term $w\in{V}$ may be relevant in some way to query $q$, it may still have a varying effect on its performance. For example, a term $w$ that is explicitly expressed in query $q$ may not be well covered in the collection $\mathcal{D}$~\cite{Buckley:2004}. As another example, a term $w$ that was implicitly derived from query $q$ (e.g., using a relevance model) may still incur a risk by including it as part of an expanded query due to possible query-drift~\cite{Buckley:2004}. Therefore, terms $w\in{V}$ may be weighed according to their potential to improve (or decline) query $q$'s performance.

For a given query $q$ and retrieved list $D_q^{[k]}$, let $P(D_q^{[k]})$ now denote the predicted quality\footnote{Such quality prediction is usually given in terms of correlation to Average Precision~\cite{CarmelBook}.} of $D_q^{[k]}$. $P(D_q^{[k]})$ may be instantiated using various post-retrieval QPP methods~\cite{CarmelBook}.
The relative influence of a term $w\in{V}$ on the performance of query $q$ (as manifested in the retrieved list $D_q^{[k]}$) is estimated by further retrieving the list $D_{q\vee{w}}^{[k]}$ ($\subseteq\mathcal{D}$). Let $P(D_{q\vee{w}}^{[k]})$ be the corresponding predicted quality of $D_{q\vee{w}}^{[k]}$.
Let $\Delta P(w;q)= P(D_{q\vee{w}}^{[k]}) - P(D_{q}^{[k]})$ denote the difference between the quality predicted for the original query $q$ and its single term expanded version $q\vee{w}$.

Finally, let $\varphi_{q}(w)\in[0,1]$ now denote the weight assigned to term $w\in{V}$, derived in this work using the following logistic (sigmoid) function\footnote{The logistic function transforms the absolute difference in predicted quality into a probabilistic [0,1] term weight representation.}:
\begin{equation}\label{relevance}
\varphi_{q}(w) =\frac{1}{1+\exp^{-\Delta P(w;q)}}.
\end{equation}

 Therefore, according to Eq.~\ref{relevance}, the larger the predicted improvement $\Delta P(w;q)$ for a given term $w\in{V}$ is, the higher will be the weight $\varphi_{q}(w)$ assigned to that term.

\section{Evaluation}\label{sec:eval}

\subsection{Proof of concept application: search re-ranking}
The proposed term weighting approach is evaluated in this work using search re-ranking as the underlying application.
Following Bendersky et al.~\cite{Bendersky:2011,Bendersky:2012}, a document $d\in D_q^{[k]}$ is re-scored based on the derived weights $\{\varphi_{q}(w)\}_{w\in{V}}$ using the following log-linear score\footnote{\textbf{TWQP} stands for \textbf{T}erm \textbf{W}eighting by \textbf{Q}uality \textbf{P}rediction.}:
\begin{equation}\label{TWQP scoring}
Score^{\sf TWQP}_{q}(d)=\sum_{w\in{V}}\varphi_{q}(w)\cdot \log\left(p^{[\mu]}_d(w)\right).
\end{equation}

Documents initially retrieved in $D_q^{[k]}$ are, therefore, re-ranked according to $Score^{\sf TWQP}_{q}(\cdot)$.

\subsection{Setup}

\paragraph*{Datasets}

\begin{table}[h]\center
\begin{tabular}{|c|c|c|c|}
  \hline
   \textbf{Corpus} & \textbf{\# of documents} & \textbf{Queries} & \textbf{Disks} \\ \hline \hline
  ROBUST & 528,155 & 301-450, 601-700 & 4\&5-\{CR\} \\ \hline
  WT10g & 1,692,096 & 451-550 & WT10g \\ \hline
  GOV2 & 25,205,179 & 701-850 & GOV2 \\ \hline
  \end{tabular}\caption{TREC data used for experiments.}\label{tab:data}
\end{table}

The TREC corpora and queries used for the evaluation are specified in
Table~\ref{tab:data}. 
Titles of TREC topics were used as queries. The Apache
Lucene\footnote{\url{http://lucene.apache.org}} open source search library (version 4.9) was used for indexing and searching documents.  Documents and queries were processed
using Lucene's English text analysis (i.e., tokenization, stemming, stopwords, etc). Lucene's implementation of the query-likelihood (QL) model with Dirichlet smoothing~\cite{Ponte:1998} was used for scoring documents. Various values of the Dirichlet-smoothing free parameter were explored $\mu\in\{100,200,\ldots,5000\}$~\cite{Ponte:1998} so as to \emph{optimize} MAP ($@1000$) (denoted \textbf{QLOpt} hereinafter).

\paragraph*{Terms extraction}
For efficiency considerations, in this work, for a given query $q$, terms in $V$ considered for weighting were induced using the RM3 pseudo relevance feedback model~\cite{Lavrenko:2001}. Usage of this relevance model allows to induce terms that are presumably relevant to the queried topic~\cite{Lavrenko:2001}.
Let $D_{q}^{[m]}$ denote the top-$m$ scored documents in $D_{q}^{[k]}$. The likelihood of a given term $w\in{V}$ according to the RM3 relevance model is calculated as follows:

\begin{equation}\label{RM3}
\begin{split}
p_{RM3}(w;D_{q}^{[m]},\mu,\lambda)= \lambda\cdot p_{q}^{[0]}(w) & \\
 + (1-\lambda)\cdot  \sum_{d\in D_{q}^{[m]}}p_{d}^{[0]}(w)&\frac{p_{d}^{[\mu]}(q)}{\sum_{d'\in D_{q}^{[m]}}p_{d'}^{[\mu]}(q)},
\end{split}
\end{equation}

where $\lambda\in[0,1]$ is a smoothing parameter, used for smoothing the query language model $p_{q}^{[0]}(\cdot)$ with the (RM1) relevance model~\cite{Lavrenko:2001}.
Finally, $V$ consists of the top-$n$ terms with the highest $p_{RM3}(w;D_{q}^{[m]},\mu,\lambda)$ likelihood. Following previous recommendations, the following parameters were fixed: $\mu=1000$, $\lambda=0.9$ and $n=100$~\cite{Lavrenko:2001}. The number of top documents $m$ in $D_{q}^{[k]}$ used for inducing the RM3 terms was further chosen as follows $m\in\{5,10,\ldots,100\}$ so as to optimize MAP ($@1000$). Let \textbf{RM3Opt} denote the induced ``optimal" RM3 model\footnote{It is worth noting that, while such simple parameter tuning was done in order to obtain the ``optimal" initial retrieval (\textbf{QLOpt}) and the ``optimal" relevance model (\textbf{RM3Opt}) that is used later on (see Section~\ref{evalmet}) for deriving a reasonable re-ranking baseline, the proposed term weighting method is completely unsupervised and requires no further tuning.}.

\paragraph*{Baseline term weighting methods}
The proposed term weighting approach, denoted \textbf{TWQP($\cdot$)} hereinafter, was compared against three other previous approaches that also utilized QPP methods for the same task~\cite{Bendersky:2008,Song:2011,Ozdemiray:2014}. The details of these methods are now shortly introduced.
\begin{itemize}
  \item \textbf{nWIG}: The normalized \emph{Weighted Information Gain} (WIG) QPP method~\cite{Zhou:2007} was used in~\cite{Bendersky:2008}, together with many other term features, for inducing term weights. For a given term $w\in V$ its \textbf{nWIG} weight is calculated as follows:
      \begin{equation}\scriptsize
      nWIG(w)=\frac{\frac{1}{m}\sum_{d\in D_{q}^{[m]}}\log p_{d}^{[\mu]}(w) - \log p_{\mathcal{D}}^{[0]}(w)}{- \log p_{\mathcal{D}}^{[0]}(w)},
      \end{equation}
       with $m=50$, following~\cite{Bendersky:2008}.
  \item \textbf{ScoreRatio}: For a given term $w\in{V}$, let $sr(w)=Score_{w}(d_1)/Score_{w}(d_k)$ denote the ratio between the score of the first and last ranked documents in $D_{q}^{[k]}$, using $q=w$ (i.e., a query with term $w$ as its single term)~\cite{Ozdemiray:2014}.
  The (sum-)normalized $sr(w)$ value is then used as the weight of term $w$. The \textbf{ScoreRatio} approach was used in~\cite{Ozdemiray:2014} for weighing query aspects for search diversification and was shown to be superior to several other state-of-the-art QPP alternatives~\cite{Ozdemiray:2014}.
  \item \textbf{SROR}: The Search Result Overlap Ratio (SROR) method proposed in~\cite{Song:2011} scores each term $w\in{\{q_i\}}$ according to the relative overlap between the initial query result and that obtained by removing term $w$ from $q$.
  The lower such overlap is, the higher the predicted query drift due to the exclusion of term $w$ from $q$ (denoted: $q-{w}$), and thus, the more important term $w$ is~\cite{Song:2011}. The \textbf{SROR} weight of a given term $w\in{\{q_i\}}$ is calculated as follows: $SROR(w)= 1-\frac{\left|D_{q}^{[k]}\cap D_{q-{w}}^{[k]}\right|}{\left|D_{q}^{[k]}\right|}$. Please note that, only terms that are explicitly specified in query $q$ may be weighed according to this method~\cite{Song:2011}.
\end{itemize}

\paragraph*{TWQP method instantiations}
To illustrate the effectiveness of the proposed \textbf{TWQP($\cdot$)} term weighting approach, three different QPP method instantiations of $P(\cdot)$ were evaluated.
Each QPP method was applied twice, once over the result of query $q$ (i.e., $D_{q}^{[k]}$) and once over the result of its expansion $q\vee{w}$ (i.e., $D_{q\vee{w}}^{[k]}$), so as to derive $\Delta P(w;q)$.
The first, denoted \textbf{TWQP(WIG)}, is based on the WIG predictor~\cite{Zhou:2007}, and is calculated as follows:
\begin{equation}\scriptsize
WIG(D_{q}^{[k]})=\frac{1}{m\sqrt{|q|}}\cdot\sum_{d\in D_{q}^{[m]}}\sum_{q_i\in{q}}\log \frac{p_{d}^{[\mu]}(q_i)}{p_{\mathcal{D}}^{[0]}(q_i)},
\end{equation}

with $m=5$, following previous recommendations~\cite{Zhou:2007}.

The second, denoted \textbf{TWQP(ScoreRatio)}, is based on the \textbf{ScoreRatio} method~\cite{Ozdemiray:2014} that was described above.
The third, denoted \textbf{TWQP(NQC)}, is based on the \emph{Normalized Query Commitment} (NQC) method~\cite{Shtok:2012}, and is further calculated as follows:
\begin{equation}\scriptsize
NQC(D_{q}^{[k]})=\frac{\sigma(D_{q}^{[m]})}{p^{[0]}_\mathcal{D}(q)},
\end{equation}

where $\sigma(D_{q}^{[m]})$ denotes the standard deviation (spread) of the score of the $m$ highest scored documents in $D_{q}^{[k]}$ (setting $m=150$, following previous recommendations~\cite{Shtok:2012}). $p^{[0]}_\mathcal{D}(q)$ further denotes the collection query likelihood~\cite{Shtok:2012}.
The higher the spread of the document scores within $D_{q}^{[m]}$, the better the performance predicted for query $q$~\cite{Shtok:2012}.

\paragraph*{Evaluation measures}\label{evalmet}
For each corpus $\mathcal{D}$ and query $q$, an initial list $D_{q}^{[k]}$ was retrieved. Each list included the top $k=1000$ ranked documents according to \textbf{QLOpt}. The RM3 terms $V$ were further used for term weighting (except for the \textbf{SROR} method, where only the initial query's terms were considered).  The top-100 documents were then re-ranked according to $Score^{\sf TWQP}_{q}(\cdot)$. To further compare against \textbf{RM3Opt}, documents in the initial list $D_{q}^{[k]}$ were re-ranked according to \textbf{RM3Opt} using the cross entropy between the (RM3) relevance
model (i.e, $p_{RM3}(\cdot,D_{q}^{[m]},\mu,\lambda)$) and their (smoothed) unigram language models (i.e., $p^{[\mu]}_d(\cdot)$, further using the same Dirichlet smoothing parameter $\mu$ derived for \textbf{QLOpt})~\cite{Lavrenko:2001}.

The effectiveness of the various term weighting approaches was evaluated using the following measures:  p@10, MAP, and MRR\footnote{The Mean Reciprocal Rank (MRR) measure captures the ability of a re-ranking method to improve the position of the first relevant document and is given by the inverse ratio of that document's position.}.
Statistically significant differences in performance were measured using the paired two-tailed t-test with a $95\%$ confidence level.

\subsection{Results}

\begin{table*}[t]
\center
\begin{tabular}{|l|
ccc|ccc|ccc|}
\hline
& \multicolumn{3}{c|}{\textbf{ROBUST}} & \multicolumn{3}{c|}{\textbf{WT10g}} & \multicolumn{3}{c|}{\textbf{GOV2}}\\\hline
\textbf{Method} & p@10 & MAP & MRR &  p@10 & MAP& MRR &  p@10 & MAP& MRR \\ \hline
\textbf{QLOpt}(init)       &43.4$^{n}$  & 25.5$^{n}$ & 66.7$^{n}$ & 29.9$^{n}$ & 20.4$^{n}$ & \textbf{56.4}$^{n}$ & 55.1 & 29.4 & 73.0   \\
\textbf{RM3Opt}~\cite{Lavrenko:2001}        & 44.5$^{in}_{s}$ & 26.0$^{in}_{s}$ & \textbf{67.3}$^{n}_{o}$ & 29.8$^{n}$ & 20.6$^{n}$ & 55.4$^{n}$  & 58.2$^{i}$ & 30.8$^{i}_{s}$ & \textbf{74.6}   \\
\textbf{nWIG}~\cite{Bendersky:2008}        & 32.0& 20.8 & 55.1 & 22.7 & 17.4 & 44.4 & 54.5 & 29.9 & 71.8  \\
\textbf{ScoreRatio}~\cite{Ozdemiray:2014}      & 45.1$^{in}$ & 26.6$^{irn}_{s}$ & 65.8$^{n}$& 31.7$^{n}$ & 21.1$^{n}$ & 53.2$^{n}$ & 60.6$^{in}$& 31.0$^{in}_{s}$ &73.4$_{\beta}$   \\
\textbf{SROR}~\cite{Song:2011}       & 43.4$^{n}$ & 25.3$^{n}$ & 65.4$^{n}$ & 30.6$^{n}$ & 20.6$^{n}$ & 57.8$^{n}$ & 58.4$^{in}$ & 30.1$^{i}$  & 75.3  \\
\hline
\textbf{TWQP(WIG}~\cite{Zhou:2007}\textbf{)}        & \textbf{46.5}$^{in}_{os}$ & \textbf{27.3}$^{irn}_{os}$ & 66.3$^{n}_{\beta}$ & \textbf{32.3}$^{n}$ & 21.9$^{n}$ & 53.2$^{n}$ & 60.9$^{in}$ & \textbf{32.0}$^{in}_{s}$ &  73.5  \\
\textbf{TWQP(ScoreRatio}~\cite{Ozdemiray:2014}\textbf{)}        &46.1$^{in}$ &\textbf{27.2}$^{irn}_{s}$  &65.2$^{n}$  & 32.1$^{n}$  & \textbf{22.1}$^{n}$ & 54.4$^{n}$ & 61.2$^{in}$ & \textbf{32.2}$^{in}_{s}$ &  71.8  \\
\textbf{TWQP(NQC}~\cite{Shtok:2012}\textbf{)}        & \textbf{46.5}$^{in}_{os}$ & \textbf{27.3}$^{irn}_{os\alpha}$ & 66.3$^{n}_{o\beta}$ & 32.2$^{n}$ & 21.9$^{n}$ & 53.4$^{n}$ & \textbf{61.3}$^{in}$ & \textbf{32.1}$^{in}_{os\alpha}$ & 73.2  \\\hline
\end{tabular}\caption{Evaluation results. \emph{i}, \emph{r}, \emph{n}, \emph{o}, \emph{s}, \emph{$\alpha$}, \emph{$\beta$} and \emph{$\gamma$} mark a statistically significant difference of a given method with the initial retrieval \textbf{QLOpt}, re-ranking according to \textbf{RM3Opt} and re-ranking according to weights derived by \textbf{nWIG}, \textbf{ScoreRatio}, \textbf{SROR}, \textbf{TWQP(WIG)}, \textbf{TWQP(ScoreRatio)} and \textbf{TWQP(NQC)}, respectively.}\label{tab:results}
\end{table*}

The results of the evaluation are depicted in Table~\ref{tab:results}. Overall, regardless of the actual QPP method that was used, \textbf{TWQP($\cdot$)} provided the most effective term weights for search re-ranking. Re-ranking according to \textbf{TWQP($\cdot$)} has provided a notable boost to the performance of the initial retrieved list. The boost in P@10 was up to +4.8\%, +6.7\% and +10.5\% more for the ROBUST, WT10g and GOV2 collections, respectively. The boost in MAP was up to +4.7\%, +2.5\% and +6.1\% more for the ROBUST, WT10g and GOV2 collections, respectively. A slight decrease, yet reasonable and insignificant, in MRR was observed.

Among the other term weighting alternatives that were evaluated, \textbf{nWIG} has exhibited a negative effect on query quality\footnote{This actually comes with no surprise, as in~\cite{Bendersky:2008} \textbf{nWIG} served only as a single feature among many others used for concept weights learning.}  while \textbf{ScoreRatio} has been the most competitive to \textbf{TWQP($\cdot$)}. Comparing \textbf{ScoreRatio} to the best \textbf{TWQP($\cdot$)} alternative on each setting, depending on the QPP method instantiation,  \textbf{TWQP($\cdot$)} has provided a better performance. Specifically, \textbf{TWQP(NQC)}, which utilizes the NQC measure as its underlying QPP method, has provided the best alternative with a significant improvement in MAP over \textbf{ScoreRatio} for the ROBUST and GOV2 datasets.

Finally, comparing  \textbf{TWQP($\cdot$)} with \textbf{RM3Opt}, it is apparent that, re-ranking according to the weights derived by \textbf{TWQP($\cdot$)} provides a more robust retrieval (in general) compared to that based on \textbf{RM3Opt} directly. By further measuring the Robustness Index\footnote{Let $\%N_{+}$ or $\%N_{-}$ be the percentage of queries in which a given method has a better or worse p@10 performance than the baseline method, respectively; then, $RI=\%N_{+}-\%N_{-}\in[-1,1]$.}~\cite{Collins-Thompson:2007b} (RI) at top-10 documents cutoff, \textbf{TWQP($\cdot$)} has provided +12.5\% (0.08 $\rightarrow$ 0.09), 0\% (no change) and +47\% (0.17 $\rightarrow$ 0.25) improvement in RI on top of \textbf{RM3Opt} for the ROBUST, WT10g and GOV2 collections, respectively.

\bibliographystyle{plain}

\end{document}